\documentstyle[11pt]{article}
\title{Revenge of the One-Family Technicolor Models}
\author{Thomas Appelquist and John Terning\\
Department of Physics, Yale University, New Haven, CT 06511}
\begin{document}
\setlength{\baselineskip}{24pt}
\maketitle
\begin{picture}(0,0)(0,0)
\put(295,240){YCTP-P9-93}
\end{picture}
\vspace{-24pt}

\begin{abstract}
\setlength{\baselineskip}{18pt}
We describe how isospin splitting and
techniquark-technilepton splitting in one-family
technicolor models can reduce the predicted value of the electroweak
radiative correction parameter $S$, without making a large
contribution to the $T$ parameter.
\end{abstract}

\section{Introduction}
Recent work \cite{Lynn,ST,uncert} has shown that the electroweak
radiative correction parameter $S$ typically receives positive
contributions in theories
where QCD-like technicolor (TC) interactions spontaneously break the
electroweak gauge symmetry. These contributions grow with the
number of technicolors
N$_{TC}$ and the number of technicolored weak doublets.  Experiments,
however,
seem to be finding $S$ to be very small or even negative
\cite{ST,Rosner}. In TC
theories with non-QCD-like dynamics, the value of $S$ could be smaller
\cite{walk}. However, it is difficult \cite{uncert}
to reliably estimate $S$ in such theories, because we cannot use
QCD as an ``analog computer".  There also exist mechanisms for
producing negative values for $S$ in certain TC models \cite{NS}.
 Thus, while it is an open question
whether there are realistic TC models that predict an acceptably
small value for $S$, previous work suggests that this can be
difficult, especially in models with more than one doublet.
In particular, one-family models, which are attractive in their
economical use of ETC gauge bosons, appear to be disfavored by the
preceding discussion.

However, previous work on estimating $S$ in models with one family of
technifermions has assumed that all the technifermions
are approximately degenerate in mass, and hence that isospin is a good
symmetry.
In this letter we point out that technifermion degeneracy is
not very likely in realistic one-family models, and furthermore that
such non-degenerate technifermions
can significantly reduce $S$, without making the
weak-isospin-violating-parameter $T$ too large.

In the next section we discuss the spectrum of technifermions in
realistic one-family models.  In section 3, we
estimate the effect that this has on $S$ and $T$. In section
4 we present our conclusions and some speculations on possible
experimental signatures for the class of TC models considered here.

\section {The Spectrum of Technifermions}

In order for a model of electroweak symmetry breaking to be
realistic, one must explain not only how the $W$'s and $Z$ get their
masses (which TC does well) but also why ordinary
fermions have such a bizarre mass spectrum.  In the TC
context this means that one must have not only a model of TC,
but also a model of extended technicolor (ETC) interactions which
feed masses down to ordinary fermions from technifermion masses.
Getting the correct masses for ordinary fermions is particularly
a problem in models with one family of technifermions if one assumes
that there is only one ETC mass
scale for each ordinary family.  Fermion masses are naively
expected to be roughly $g_{ETC}^2 4\pi f^3/M^2$, where $f$ is the
Nambu-Goldstone boson (NGB) decay constant, $M$ is the
mass of the ETC gauge boson, and $g_{ETC}$ the ETC gauge coupling.
It is then difficult to see how one can arrange for
the $t$ quark to have a mass around $150$ GeV, while the $\tau$
lepton has a mass of $1.8$ GeV, when both masses arise through
the exchange of the same ETC gauge boson.

A possibility is that QCD
interactions in concert with a near-critical ETC interaction can
greatly enhance the masses of quarks over leptons \cite{color1,color2}.
One calculation \cite{color2}
found that quark masses could be up to two orders of magnitude larger
than lepton masses, without excessive fine-tuning of the ETC
interaction. The same would then be true for techniquark and
technilepton
masses renormalized at the ETC scale. At TeV energies and below,
this QCD enhancement also makes the techniquarks
($U$,$D$) heavier
than the technileptons ($N$,$E$), but by a much smaller factor, say
of order 3-5.
We note that the bulk of the $W$ and $Z$ masses in such a
model would come from the techniquarks, since, for example,
the mass of the $Z$ is given to lowest order by
\begin{equation}
  M_Z^2 = {{g^2+g^{\prime 2}}\over 4}
\left({1\over 2}f_N^2+{1\over 2}f_E^2+3 f_Q^2\right) ~,
\label{M_Z}
\end{equation}
where $f_N$, $f_E$, and $f_Q$ are the NGB decay constants
associated with NGB's composed of technineutrinos, technielectrons,
and techniquarks
respectively. Here $g$ and $g^{\prime}$ are the $SU(2)_L$ and $U(1)_Y$
gauge couplings.
We expect the NGB decay constants to have ratios similar to
corresponding ratios of TeV-scale technifermion masses.

Since $T$ is small, one must require
in such a model (with the techniquarks dominating the weak-scale
physics) that weak isospin symmetry is not broken too badly for the
techniquarks at TeV scales and below. This can be difficult with the
very different $t$ and $b$ quark masses which must be arranged for in
the theory, since the $U$ and $D$ masses at the ETC scale are typically on
the order of the $t$ and $b$ masses. One way to arrange the $t$-$b$
hierarchy, and yet to keep the $U$ and $D$ masses close to each other
at  TeV scales and above, is to have different ETC scales for the $t$
and $b$
\cite{Einhorn}. Another way is through extra mixing for the $b$
(e.g. a generalization of the scenario
in reference \cite{Simmons}).  In this paper, it will be assumed that
some mechanism of this sort leads to a relatively small splitting
between the TeV-scale $U$ and $D$ masses.

Since the lighter technileptons make only a small
contribution to the gauge boson masses, however, the technielectrons
and technineutrinos are allowed to have
substantially different TeV-scale masses.
That they should be
different is natural in an ETC theory, which
must explain the large splitting between neutrinos
and charged leptons.  However one arranges for this,
it must involve different ETC
couplings for the technielectron and technineutrino.  If the
ETC interactions are near-critical
(in order to get a large $t$-$\tau$ splitting) then they
can have a potentially large effect, and we expect that the
technielectron and the
technineutrino will have significantly different masses. In what
follows we will assume that, as in the
ordinary lepton pattern, the technineutrino will be lighter than the
technielectron.

Thus we expect that in a realistic one-family model (without a
plethora of ETC scales), there will
be a hierarchy of technifermion masses. At TeV scales and below, we
expect to have heavy techniquarks which are approximately degenerate,
a much lighter technielectron,
and an even lighter technineutrino. If the bulk of the $W$ and $Z$
masses is to come from the techniquarks through equation 1, then using
a naive scaling from QCD, the TeV-scale
masses of the $U$ and $D$ can be estimated to be approximately
$860/\sqrt{N_{TC}}$ GeV.
For purposes of our numerical estimates, we will use $N_{TC}=2$ (which
minimizes $S$).   The main constraint on
the technilepton masses
is that they must be larger than roughly $M_Z/2$.  For our estimates
we will
take the mass of the technielectron $E$ to be $150$ GeV and the
mass of the technineutrino $N$ to lie in the range $50$ to $150$ GeV.

It should
be pointed out that this pattern of mass scales is different
from that envisaged in conventional one-family models.
The intrinsic scale of TC (which we take to be around 100 GeV) is
smaller than is usually considered,
since techniquarks receive a large part of their masses from
near-critical
ETC (together with QCD) interactions.  We are thus assuming that ETC
interactions have a major effect on the dynamical, TeV-scale
technifermion masses, rather than being a small perturbation.
In the next section we will estimate the effect that this
unusual spectrum of technifermions has on the $S$ and $T$
parameters.

\section{Precision Electroweak Measurements}

We first  estimate the value of $S$ in realistic
one-family technicolor models, as described above. The $S$ parameter
corresponds to a certain term in the chiral
Lagrangian description of the electroweak interactions
\cite{ST,Longhitano}. It is generated by integrating out everything
except the standard model corrections themselves. This will include
contributions from the pseudo-Nambu-Goldstone
bosons (PNGB's), referred to here as the ``low-energy" contributions,
as well as ``high-energy" contributions from the techniquarks and
technileptons.

In order to calculate the PNGB contribution, we must first
estimate the spectrum of PNGB's.
When discussing one-family TC models, it is often
assumed that the approximate global chiral
symmetry is
$SU(8)_L \otimes SU(8)_R \otimes U(1)_V$  (corresponding to $3$
techniquark doublets, and one
technilepton doublet)\footnote{It is assumed here that
TC and/or the near-critical ETC dynamics distinguishes between
technifermions and anti-technifermions \cite{vac}. For the $SU(2)$ TC
group to be
employed in our numerical estimates, it is the ETC interaction that
must provide this distinction \cite{Sundrum}.}. The large splitting in
technifermion masses
discussed in the last section indicates that in the type of models we
are discussing, the
approximate global chiral symmetry of one-family of technifermions is
rather\footnote{The $6$ corresponds to the techniquarks,
and the $2$ to the technileptons.}
$SU(6)_L \otimes SU(6)_R \otimes SU(2)_L \otimes U(1)_{2R}
\otimes U(1)_{8L} \otimes U(1)_{8R}
\otimes U(1)_V$. The $U(1)_{8L}$ and $U(1)_{8R}$
correspond to the generators of $SU(8)_L$ and $SU(8)_R$
which are proportional to ${\rm diag}(1,1,1,1,1,1,-3,-3)$, and
$U(1)_{2R}$ corresponds to the diagonal generator of $SU(2)_R$.
TC interactions spontaneously break this global chiral symmetry
to $SU(6)_V \otimes U(1)_{2V} \otimes U(1)_{8V} \otimes U(1)_V$.
Thus instead of
having $60$ PNGB's as is usually assumed, we have only $36$.
The explicit breaking of $SU(8)_{L} \otimes SU(8)_{R}$ is so large
here that the color triplet PNGB's usually present in one-family
models are not expected to exist.

The PNGB's and NGB's can thus be enumerated as follows (we display
their quantum numbers in terms of technifermion fields):
\begin{eqnarray}
\Theta_a^\alpha & \sim & {\overline Q} \gamma_5 \lambda_a
\tau^\alpha Q ~,\nonumber \\
\Theta_a & \sim & {\overline Q} \gamma_5 \lambda_a Q ~, \nonumber\\
P_Q^\pm & \sim & {\overline Q} \gamma_5
\tau^\pm Q ~, \nonumber \\
P_Q^3 & \sim & {\overline Q} \gamma_5
\tau^3 Q  ~, \nonumber \\
P_L^\pm & \sim & {\overline L} \;\frac{1}{2} (1-\gamma_5)
\tau^\pm L ~, \\
P_L^3 & \sim & {\overline L} \gamma_5
\tau^3 L  ~, \nonumber \\
P^0 & \sim & {\overline Q} \gamma_5 Q - 3 {\overline L} \gamma_5 L
 ~, \nonumber
\label{PNGB's}
\end{eqnarray}
where $Q$ represents the techniquarks, $L$ the technileptons,
the $\lambda_a$'s are $SU(3)_C$ generators, and the
$\tau^\alpha$'s are Pauli matrices.  The NGB's which are eaten by
the $W$'s and $Z$ are linear combinations of the $P$'s. The PNGB
mass eigenstates are formed from the orthogonal combinations (i.e.
the coupling to an electroweak gauge boson vanishes).
In general there is mixing between the $P^3$'s and the
$P^0$, which is model dependent.  In the limit of large isospin
splitting we expect that the mass eigenstates will be approximately
$P_N \approx {\overline N} \gamma_5 N$ and  $P_E \approx {\overline E}
\gamma_5  E$, with
a small admixture of techniquarks.

The PNGB
contribution to $S$ comes
from loops of the $\Theta^\alpha_a$'s and the $P^\pm$'s. It is given by
\cite{ST,RP}:
\begin{equation}
S_{PNGB}={1\over{6 \pi}}\left[ \ln\left({\Lambda_\chi\over
{M_{P^\pm}}}\right)
 + 8 \ln\left({ \Lambda_\chi\over M_{\Theta^\alpha_a}}\right)\right]~,
\label{SPNGB}
\end{equation}
where $\Lambda_\chi$ is the ultraviolet cutoff scale in the loop
integration. We take $\Lambda_\chi$ to be the scale where $SU(6)_L
\otimes SU(6)_R$ chiral perturbation
theory breaks down, which is roughly $4 \pi f_Q/\sqrt{6}\approx 720$
GeV.  Using this cutoff probably overestimates the
contribution from the $P^\pm$ loop, since these PNGB's are mainly
composed of the (lighter) technileptons, and thus should be associated
with a smaller decay constant, and hence a lower cutoff.
The mass
of the $P^\pm$ is very model dependent since it arises mainly through
ETC interactions. This means that the squared  mass is proportional to
technifermion condensates, and is thus sensitive to details
of the TC dynamics, e.g. whether the TC coupling is running or walking.
Experimentally we know that the $P^\pm$ must be at least as
heavy as $\approx M_Z/2$.  We will take the range to be
\begin{equation}
50 {\rm GeV} < M_{P^\pm} <150 {\rm GeV} ~.
\label{Mppm}
\end{equation}
Fortunately, since the $P^\pm$ makes only a small
contribution to $S$, our final results are not that sensitive to
this uncertainty.  The mass of the $\Theta^\alpha_a$'s has been
estimated in QCD-like TC theories \cite{vac} to be
$245$-$315$ GeV.  This estimate relies on scaling up a QCD
dispersion relation.  If the TC dynamics are not QCD-like, then
this mass estimate will be modified.  We will consider the range
\begin{equation}
250{\rm GeV} < M_{\Theta^\alpha_a} <500 {\rm GeV}~.
\end{equation}
With this range of PNGB masses, we find $0.2 < S_{PNGB} < 0.6$.

The calculation of the ``high-energy" contribution to $S$ is more
difficult, since it directly involves non-perturbative physics.  The
two methods used in the past \cite{ST,walk}
(scaled-up QCD data from dispersion relations or chiral
Lagrangians, or
non-local chiral quark models) have relied on the assumption that
isospin is not broken.  Here we are assuming that isospin is badly
broken for technileptons.  The non-local
chiral quark model could be generalized
to overcome this difficulty, but not without considerable
effort.   Even a modified dispersion relation approach would not be
straightforward, since the spectrum is unlike that of QCD.

A  naive approach, neglecting strong technicolor interactions, will be
adopted here.  We note that
in the case of one-doublet\footnote{Where there is no PNGB
contribution to $S$.}, QCD-like TC theories, both methods
mentioned above arrive at a value
for the ``high-energy" contribution to $S$ that is about twice as
large as the perturbative, one-technifermion loop
estimate (using  constant technifermion masses).  We also note that
studies of walking \cite{walk} TC arrive at values
of $S$ that are as small as half the scaled-up QCD result, i.e.
approximately equal to the perturbative result (using constant masses).
On the other hand, in the case
of one-family QCD-like theories (with  $N_{TC}=3$) the
dispersion-relation
result for the ``high-energy" piece (i.e. $S-S_{PNGB}$)  is about
half as big as the perturbative result.  Here we will
simply calculate the perturbative (one-technifermion loop)
contribution,
using twice and one half of this value to estimate the range of
possible values for the ``high-energy" contribution.

With this assumption, the calculation of the ``high-energy" techniquark
contribution to $S$ is straightforward.  We use the definition
\cite{ST}
\begin{equation}
S_{TQ}\equiv- 8 \pi \Pi^{TQ \prime}_{3Y}(q^2=0) ~,
\label{TQ}
\end{equation}
where
\begin{equation}
\Pi^{TQ}_{3Y}=\left(q_U - q_D \right) \Pi^{TQ}_{LR} ~,
\label{3Y,TQ}
\end{equation}
$q_U$ and $q_D$ are the electromagnetic charges of the $U$ and $D$, and
the prime indicates a derivative with respect to $q^2$. It has
been assumed here that isospin is a good approximate symmetry for
techniquarks.
The $\Pi$'s refer to the coefficients of $i g_{\mu \nu}$ in vacuum
polarizations (with gauge couplings factored out, as usual),
and $L$ and $R$ refer to left- and
right-handed currents. Using constant masses for the perturbative
calculation leads to the standard result \cite{ST}:
\begin{equation}
S_{TQ}={{N_{TC} N_C}\over{6 \pi}}~.
\label{STQ}
\end{equation}
With $N_{TC}=2$, our estimated range is therefore $0.2 < S_{TQ} < 0.6$.

We turn next to the calculation of the contribution to $S$ from
technileptons.  We will,
of course, {\em not} assume that isospin is a good approximate
symmetry for the technileptons.  We note that the
technilepton masses being employed here are too small
for the original\footnote{i.e. keeping only the leading term in a
Taylor
series expansion of the vacuum polarization} definition of $S$ to be
justified. The contribution
to $S$ from the technileptons is therefore defined as
(cf. ref.~\cite{Mar})
\begin{equation}
S_{TL}\equiv - 8 \pi {{\Pi^{TL}_{3Y}(q^2=M_Z^2)-\Pi^{TL}_{3Y}(q^2=0)}
\over{M_Z^2}} ~.
\label{TL}
\end{equation}
We note that equation (\ref{TL}) reduces to the same form as
equation (\ref{TQ}) when the masses of the technileptons become
much larger than $M_Z$.  Now,
\begin{equation}
\Pi^{TL}_{3Y}={Y \over 2} \left(\Pi^N_{LL}-\Pi^E_{LL} \right)
+q_N \Pi^N_{LR} - q_E \Pi^E_{LR} ~.
\label{3Y,TL}
\end{equation}
  As we will see, it is the first term in
equation (\ref{3Y,TL}) that can give a negative contribution to $S$.

The required one-loop results (with constant masses) are well
known \cite{Peskin}:
\begin{eqnarray}
\Pi_{LL}(m_1,m_2,q^2)&=&{{-1}\over {4 \pi^2}} \int_0^1 dx \ln\left(
{{\Lambda^2}\over{m^2-x(1-x)q^2}}\right)
\left(x(1-x)q^2-{1 \over 2}m^2\right) ~,\nonumber\\
\Pi_{LR}(m_1,m_2,q^2)&=&{{-m_1 m_2}\over {8 \pi^2}} \int_0^1 dx
\ln\left({{\Lambda^2}\over{m^2-x(1-x)q^2}}\right) ~,
\label{Pi's}
\end{eqnarray}
where $\Lambda$ is an ultraviolet cutoff, $m_1$ and $m_2$ are the
masses of the fermions in the loop, and $m^2=x m_1^2+(1-x)m_2^2$.
This leads to:
\begin{eqnarray}
S_{TL}&=&{{-N_{TC}}\over{\pi}}\int_0^1 dx
\ln\left({{M_E^2-x(1-x)M_Z^2}
\over{M_N^2-x(1-x)M_Z^2}}\right)x(1-x)\nonumber \\
& &+{{N_{TC}}\over{2\pi}}{{M_N^2}\over{M_Z^2}}
\int_0^1 dx \ln\left({{M_N^2}
\over{M_N^2-x(1-x)M_Z^2}}\right) \nonumber\\
& &+{{N_{TC}}\over{2\pi}}{{M_E^2}\over{M_Z^2}}
\int_0^1 dx \ln\left({{M_E^2}
\over{M_E^2-x(1-x)M_Z^2}}\right)
\label{STL}
\end{eqnarray}
Thus we find that the technileptons can give a negative contribution
to $S$.  For example with $N_{TC} = 2$, $M_E = 150$ GeV, and
$M_N = 50$ GeV,
we obtain $S_{TL}=-0.2$, and thus an estimated range of $-0.1$ to
$-0.4$.

Putting all three contributions (equations (\ref{SPNGB}), (\ref{STQ}),
(\ref{STL})) together,
we arrive at the
estimates given in Table 1, for the smallest TC group: $SU(2)_{TC}$.
Recent fits to experimental data \cite{Rosner} (with a $t$ quark
mass of $140$ GeV, and translating\footnote{Since TC is an
alternative
to the standard model Higgs sector, one must subtract off from $S$ the
Higgs
contribution to vacuum polarizations which is already included in
standard model fits to
data, so the value of $S$ depends on the Higgs mass used in a given
fit.} to $M_{\rm Higgs}=\Lambda_\chi$)
lead to upper limits on
$S$ (at the 90\% confidence level) that are typically no more than
a few tenths.  The reduction in the theoretical
prediction discussed here could therefore be important in attaining
agreement with experiment.
\begin{table}[htb]
\begin{center}
\begin{tabular}{|l||c||c||c|}\hline\hline
$M_N$ &  $S_{TQ}$&$S_{TL}$  & $S$ \\ \hline\hline
$50$ GeV & $0.2$ - $0.6$& $-0.1$ - $-0.4$& $-0.02$ -
$1.1$ \\ \hline
$100$ GeV  &$0.2$ - $0.6$ & $0.01$ - $0.03$ & $0.4$ -
 $1.3$ \\ \hline
$150$ GeV  &$0.2$ - $0.6$ &$0.06$ - $0.2$ & $0.5$ -
 $1.5$ \\ \hline\hline
\end{tabular}
\end{center}
\caption{Estimates of $S=S_{PNGB}+S_{TQ}+S_{TL}$, for different values
of the technineutrino mass
($M_N$), with $M_E = 150$ GeV, and $N_{TC}=2$.  We have used $0.2 <
S_{PNGB} < 0.6$.}
\end{table}

It can be seen from equation (\ref{3Y,TL}) that if we were doing a
more realistic calculation, taking into account the
strong interaction dynamics of the technifermions, it is the
techni-$\rho$
composed of technineutrinos (which would be lighter than the
techni-$\rho$
composed of technielectrons) that gives a negative contribution to
$S$.  If techni-$\rho$'s are lighter than standard estimates suggest
\cite{uncert}, then the negative contribution to $S$ will be enhanced.

We next discuss the computation of $T$ in realistic
one-family TC models, arising
from isospin splitting in PNGB's and technileptons.
The point of the estimate is to show that, although
there is a large isospin breaking for technileptons, this does
not lead to a large contribution to $T$. Since the bulk of the $W$ and
$Z$ masses come from the heavier techniquarks, the large isospin
breaking in the (relatively light) technileptons gives a much smaller
contribution than if the
technileptons were the sole contributors to the gauge boson masses.

We again use one-technifermion-loop graphs (with constant masses) to
estimate  the ``high-energy" contribution, and chiral perturbation
theory to estimate the ``low-energy" contribution.
First recall that \cite{ST}
\begin{equation}
\alpha T \equiv \Delta \rho_* \equiv {{g^2+ g^{\prime 2}}
\over{ M_Z^2}} \left[\Pi_{11}(0)-\Pi_{33}(0)\right]~.
\end{equation}

We first consider the ``high-energy" technilepton contribution to $T$.
The perturbative result for one fermion loop is \cite{T}:
\begin{equation}
\alpha T_{TL}={{\left(g^2+ g^{\prime 2}\right) N_{TC}}
\over{64 \pi^2  M_Z^2}}
\left[M_N^2 +M_E^2 -{{4 M_N^2 M_E^2}\over{M_N^2-M_E^2}}
\ln\left({M_N\over M_E}\right)\right]~.
\end{equation}
Thus for $N_{TC}=2$, $M_E = 150$ GeV, and $M_N= 50$
GeV, we find $\Delta\rho_{TL}=\alpha T_{TL} = 0.26\%$. We take
our estimated range to be from one half of to twice this value:
$0.1\%< \Delta\rho_{TL} <0.5\%$. If the
isospin splitting is smaller, then these numbers become even smaller.
For $M_N= 100$, for example, we find $0.03\%<\Delta\rho_{TL}<
0.1\%$.

The contribution to $T$ from PNGB's is zero \cite{RP}, unless there is
mass splitting within isospin multiplets.  Since there is a large
isospin splitting for technileptons, we should expect some
contribution from the PNGB's composed of technileptons.  We assume
that the PNGB
isospin eigenstate ($I=1$, $I_3=0$) is given by the linear
combination of mass eigenstates $c_\theta P_N-s_\theta P_E$. For
maximal isospin breaking, $c_\theta = s_\theta =
1/\sqrt{2}$. Using the
results in ref.~\cite{RP}, we find
\newpage
\begin{eqnarray}
\lefteqn{\alpha T_{PNGB} = }\nonumber \\
& &{{g^2+ g^{\prime 2}} \over {32 \pi^2 M_Z^2}}
\left[\begin{array}{l}
\displaystyle c^2_\theta\int_0^1 dy \, \Delta_N
\ln\left({{\Lambda_\chi^2}\over{\Delta_N}}\right)
\displaystyle +s^2_\theta\int_0^1 dy \, \Delta_E
\ln\left({{\Lambda_\chi^2}\over{\Delta_E}}\right)  \\
\displaystyle - M_{P^\pm}^2
\ln\left({{\Lambda_\chi^2}\over{M_{P^\pm}^2}}\right)
\end{array}\right],
\end{eqnarray}
where
\begin{eqnarray}
\Delta_N &=& M_{P_N}^2 + (1-y)(M_{P^\pm}^2 -M_{P_N}^2)~, \nonumber\\
\Delta_E &=& M_{P_E}^2 + (1-y)(M_{P^\pm}^2 -M_{P_E}^2)~.
\end{eqnarray}
We will examine the plausible and broad range of masses given by
equation (\ref{Mppm}), and by
\begin{eqnarray}
M_{P^\pm}^2 & < M_{P_E}^2< & 2 M_{P^\pm}^2~, \\
10 {\rm GeV} & < M_{P_N} < & M_{P^\pm}~.
\end{eqnarray}
Taking $c_\theta = s_\theta=1/\sqrt{2}$, we find
that the PNGB contribution to $\Delta \rho_*$
is $-0.3\% <\Delta \rho_{PNGB}<0.2\%$. Thus the contribution to
$\Delta\rho_*\equiv \alpha T$ from technileptons and PNGB's (for the
parameters we
have considered above) is in the following range:
\begin{equation}
-0.3\% < \Delta\rho_{TL}+\Delta\rho_{PNGB} <0.7\% ~.
\end{equation}
Recent global fits to the data \cite{Rosner},
show that most of the above range
is experimentally allowed.

\section{Conclusions}
We have argued that in realistic one-family
TC models the techniquarks will be much heavier than technielectrons,
which in turn will be much heavier than technineutrinos.  We have
estimated the possible effects on
precision electroweak measurements that arise in TC models of this
type, and noted that $S$ can be substantially smaller than traditional
estimates suggest.

We note that if technineutrinos are really as light as we
have been considering in this letter, then the techni-$\rho$ composed
of technineutrinos ($\rho_N$) will also be light, presumably in the
range $100$-$300$ GeV. Such a particle could provide a
spectacular signal at LEP II or, if it is somewhat heavier, at the next
$e^+ e^-$ collider or the SSC. This would be the first experimental
signature of
the type of model being considered here. We expect the following
$\rho_N$ decay modes (in order of predominance, if kinematically
allowed):  $P_N$ pairs (which in
turn decay into third generation fermions),  $P_N Z$ , $P^\pm W^\mp$,
 $W^\pm W^\mp$, and  $ZZ$
(cf. ref~\cite{Lane}).  If the $\rho_N$ is
too light for the possibilities listed above, then it
will be extremely narrow, and decay predominantly into quarks and
leptons through the $Z$, and also (with a small branching fraction)
into third-generation fermions through an ETC gauge boson.

\noindent\medskip\centerline{\bf Acknowledgments}
\vskip 0.15 truein
We would like to thank R. Sundrum for helpful conversations, and for
a critical reading of the manuscript.
This work was supported in part by the Texas National Research
Laboratory
Commission, and by the  Department of Energy under contract
\#DE-AC02ERU3075.

\end{document}